\documentstyle[twoside,fleqn,espcrc2,epsf]{article}
 
\newcommand{\AmS}{{\protect\the\textfont2
  A\kern-.1667em\lower.5ex\hbox{M}\kern-.125emS}}
 
\title{\hfill{\normalsize SLAC-PUB-7274}\\[-0.2cm]
       \hfill{\normalsize NORDITA 96/52-P}\\[-0.2cm]
       \hfill{\normalsize hep-ph/9609266}\\[-0.2cm]
       \hfill{\normalsize September 1996}\\[0.5cm]
On the infrared sensitivity of the longitudinal cross section in 
$e^+e^-$ annihilation\thanks{To appear in the Proceedings of the 
High-Energy Physics International Euroconference 
on Quantum Chromodynamics (QCD96), Montpellier, France, 
4-12 July 1996}}
\author{
M. Beneke\address{
SLAC, P.O.~Box 4349, Stanford, CA~94309, U.S.A.}
\thanks{ Research supported by the Department of Energy under 
contract DE-AC03-76SF00515.},
    V.M. Braun$^{\rm b}$ and L. Magnea\address{
NORDITA, Blegdamsvej 17, DK--2100 Copenhagen, Denmark}\thanks{On 
leave of absence from 
Universit\`a di Torino, Italy}
     }                     
\begin{document}
 
\begin{abstract}
We have calculated the contributions proportional to 
$\beta_0^n\alpha_s^{n+1}$ to the longitudinal fragmentation 
function in $e^+e^-$ annihilation to all orders of perturbation 
theory. We use this result to estimate  
higher-order perturbative corrections and nonperturbative
corrections to the longitudinal cross section $\sigma_L$ and 
discuss the prospects of determining $\alpha_s$ from $\sigma_L$. 
The structure of infrared renormalons in the perturbative expansion
suggests that the longitudinal cross section for hadron production 
with fixed momentum fraction $x$ receives nonperturbative 
contributions of order $1/(x^2Q^2)$, whereas the total cross 
section has a larger, $1/Q$ correction. This correction arises 
from very large longitudinal distances and is related to 
the behaviour of the Borel integral for the cross section 
with fixed $x$ at large values of the Borel parameter. 
\end{abstract}
 
\maketitle
\section{Introduction}
 
The ALEPH \cite{ALEPH} and OPAL \cite{OPAL} collaborations have
measured the dependence of single-particle 
inclusive cross sections in $e^+e^-$ annihilation on the scattering
angle $\theta$ between the observed hadron $h$ and the incoming electron 
beam. The angular dependence discriminates between 
contributions from transversely and longitudinally polarized 
virtual bosons, and from $Z^0$-photon interference \cite{NAS94} 
\begin{eqnarray}\label{def:TL}
\lefteqn{
\frac{d^2\sigma^h}{dx d\cos\theta}(e^+e^-\to hX) =}
\nonumber\\
&\hspace*{-0.5cm} =&
{}\!\!\!\!\!\!\frac{3}{8}(1+\cos^2\theta) \frac{d\sigma_T^h}{dx}(x,Q^2)
+\frac{3}{4}\sin^2\theta\, \frac{d\sigma_L^h}{dx}(x,Q^2)
\nonumber\\&&{}\!\!\!\!\!\!+
\frac{3}{4}\cos\theta\, \frac{d\sigma_A^h}{dx}(x,Q^2) .
\end{eqnarray}
In the following, dropping the superscript `$h$' implies 
summation over all hadrons $h$. 

In this paper we concentrate on the longitudinal cross section. 
It is given as a convolution
of a parton fragmentation function $D^h_p$ ($p=q,\bar{q},g$)
with a partonic cross section ${d\hat\sigma_L^p}/{dx}$
\begin{equation}
\frac{d\sigma_L^h}{dx}(x,Q^2) =\sum_p\int_x^1\frac{dz}{z} 
\frac{d\hat\sigma_L^p}{dz}(z) D_p^h(x/z,Q^2) .
\end{equation} 
The perturbative expansion of 
the longitudinal parton cross section starts at order 
$\alpha_s$.

Summed over all hadrons, the fragmentation functions
satisfy the energy conservation 
sum rule $\sum_h \int_0^1\!dx\,x D^h_p(x,Q^2)=1$.  
Consequently, the integrated longitudinal cross section
is an infrared (IR) safe quantity which is calculable in perturbation
theory
\begin{eqnarray}\label{sigmaL}
\sigma_L &\equiv & \sum_h\frac{1}{2}\int_0^1\!\! 
dx \, x\frac{d\sigma_L^h}{dx}
= \sum_p\frac{1}{2}\int_0^1\!\! dx \,x \frac{d\hat\sigma_L^p}{dx}
\nonumber\\
&&\hspace*{-1.4cm}=\,\sigma_0\!\left[
\frac{\alpha_s}{\pi}+(14.583-1.028N_f)\left(\frac{\alpha_s}{\pi}\right)^2
\!+\ldots\right] .
\end{eqnarray}
Here $\sigma_0$ is the Born total $e^+e^-$ annihilation cross section,
$\alpha_s\equiv \alpha_s(Q)$ and $N_f$ is the number of active fermion
flavours. The next-to-leading order contribution has been obtained 
in \cite{RIJ96}. OPAL \cite{OPAL} has measured $\sigma_L$ the $Z_0$ peak: 
\begin{equation}
  \sigma_L/\sigma_{\rm tot}(M_Z^2) =0.057\pm 0.005~.
\end{equation}

One of the main motivations for the present study is 
to investigate whether 
measurements of the total longitudinal cross sections can 
yield a precise determination of the strong coupling. 
This requires that we control higher-order perturbative 
corrections and nonperturbative effects, both of which are 
expected to be much larger for $\sigma_L$ than for the 
total cross section $\sigma_{tot}$. We address both types of 
corrections in this report, by
studying the structure of IR renormalons, a certain 
class of higher-order perturbative corrections, for the longitudinal 
cross section. 

The study of nonperturbative effects in fragmentation functions
is an interesting topic in its own right \cite{BBM}. 
The light-cone expansions for fragmentation functions and for structure 
functions in DIS are similar \cite{BB91}, and suggest   
that nonperturbative effects in both cases are of order 
$1/Q^2$ and can be described in terms of multi-parton distributions.
In contrast to DIS, however, the relevant operator structures for 
fragmentation are essentially nonlocal and cannot be expanded at small 
distances. Hence the usual operator product expansion does not apply
and the status of the light-cone expansion 
is less established. Hadronization models generically introduce 
nonperturbative corrections of order $1/Q$, while
current data on scaling violations in fragmentation 
do not distinguish between $1/Q$ or $1/Q^2$ behaviour. 
A nonperturbative correction of order $1/Q$ 
to the total longitudinal cross section was suggested in \cite{WEB94}
as a consequence of phase-space reduction in the one-loop diagram 
calculated with a massive gluon. We address these apparently conflicting 
statements below.

While this work was in writing, Dasgupta and Webber \cite{DAS96} 
have addressed a similar set of questions with closely related methods.

\section{General formalism}

There is suggestive evidence from exact low-order results that 
$\beta_0$ is a large parameter and that keeping corrections of 
order $(\beta_0\alpha_s)^n$ in higher orders resums important 
contributions. Moreover, the infrared renormalons encoded in 
the corresponding series can elucidate the power-behaviour of 
nonperturbative corrections and, perhaps, even their $x$-dependence, as 
in the case of $d\sigma_L/d x$. The $(\beta_0\alpha_s)^n$ corrections 
can be traced by inserting a chain of fermion loops  
into the gluon propagator, and by restoring the full QCD $\beta$-function 
coefficient $\beta_0 =-1/(4\pi)[11-2/3N_f]$ from 
the dependence on $N_f$. For $\sigma_L$ we 
obtain two contributions, according to whether the registered 
parton comes from the primary vertex or a fermion loop. The 
corresponding diagrams are shown in Fig.~1 and Fig.~2, for the 
contributions of the `primary' and `secondary' quarks, respectively. 
Note that the secondary quark contribution reduces to the gluon 
contribution at lowest order in $\alpha_s$.  
\setlength{\unitlength}{0.7mm}
\begin{figure}[t]
\vspace{2.3cm}
\hspace*{-3cm}
\begin{picture}(120,200)(0,1)
\mbox{\epsfxsize13.0cm\epsffile{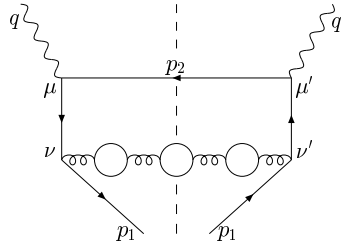}}
\end{picture}
\vspace*{-13.0cm}
\caption{The `primary' quark contribution to $\sigma_L$. Sum over
all possible insertions of the bubble chain is understood.}
\end{figure}
\setlength{\unitlength}{0.7mm}
\begin{figure}[t]
\vspace{1.3cm}
\hspace*{-3cm}
\begin{picture}(120,200)(0,1)
\mbox{\epsfxsize13.0cm\epsffile{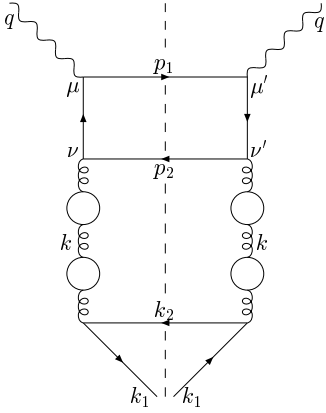}}
\end{picture}
\vspace*{-11.0cm}
\caption{The `secondary' quark contribution to $\sigma_L$. Sum over
all possible insertions of the bubble chain is understood.}
\end{figure}

The evaluation of the two classes of diagrams, for an arbitrary 
number of internal fermion loops, and for their sum,
is relatively straightforward by means of the dispersion technique
developed in \cite{BEN95,NEU95,BBB}, in terms of the distribution function 
over the invariant mass $k^2$ of the bubble chain:
\begin{eqnarray}\label{xi-distributions}
\lefteqn{
\frac{d\hat\sigma_L^{\rm{[p,s]}}}{dx}(x,\xi=k^2/Q^2)\equiv  
        }
\nonumber\\
&\equiv & \frac{8\pi}{N_c q^2}\int\! d\,{\rm Lips}[p_1,p_2,k]\,
\frac{1}{k^2}{\cal M}_{\mu\nu}{\cal M}_{\mu'\nu'}^\ast
\nonumber\\
&&\hspace*{-0.4cm}{}\times 
 \int\! d\,{\rm Lips}[k_1,k_2]\,
\frac{1}{2}{\rm Tr}[\gamma_\nu\!\not\!k_1\gamma_{\nu'}\!\not\!k_2]
\nonumber\\
&&\hspace*{-0.4cm}{}\times \left\{\begin{array}{l}
 \frac{p_{1,\mu}p_{1,\mu'}}{2|\vec p_1|^2}
 \delta\left(x-\frac{2p_1 q}{q^2}\right)~~{\mbox{`primary'}}
                                \\
 \frac{k_{1,\mu}k_{1,\mu'}}{2|\vec k_1|^2}
 \delta\left(x-\frac{2k_1 q}{q^2}\right)~~{\mbox{`secondary'}}
\end{array}\right .
\end{eqnarray}
The notation for momenta and Lorenz indices corresponds to Fig.~2.
We denote by $[p]$ ($[s]$) the contribution of the `primary' (`secondary')
quark, while $\int\!d\,{\rm Lips[\ldots]}$ are Lorentz-invariant phase 
space integrals with the momentum conservation $\delta$-function included, 
and ${\cal M}_{\mu\nu}{\cal M}_{\mu'\nu'}^\ast$ is the matrix element 
for the primary $q\bar{q} g$ amplitude squared.

Note that in the case of the `primary' quark contribution the phase space
integral over $k_1,k_2$ is proportional to $k^2$, so that the result takes 
the form of the one-loop diagram calculated with a gluon of mass $k^2$. 
This equivalence does not hold for the registered `secondary' quark
because of the nontrivial longitudinal projector. Because of this 
inequivalence, the restoration of $\beta_0$ from fermion loops 
is not unambiguous and the relation of fermion loop chains with 
running coupling effects is partially lost. In practice, we have 
found that the numerical differences are small, so that a detailed 
discussion is deferred to \cite{BBM}. The analytic expressions for 
the primary and secondary quark contribution are rather lengthy and 
will also be given there.

Given the invariant mass distributions (in $\xi=k^2/Q^2$), 
finite order results 
are obtained in terms of the logarithmic integrals \cite{BEN95,BBB}
\begin{equation}
\int_0^1 d\xi \,\ln^n\xi \,\frac{d}{d\xi}\frac{d\hat\sigma_L}{dx}(x,\xi) . 
\end{equation} 
The sum of the series, defined by a principal value prescription for 
the Borel integral, equals 
\begin{eqnarray}\label{BS}
\frac{d\hat\sigma_L^{(BS)}}{dx} &=&
\int_0^1\!d\xi \,\Phi(\xi)\,\frac{d}{d\xi}\frac{d\hat\sigma_L}{dx}(x,\xi) 
\nonumber\\
&&\hspace*{-1cm}{}+\left[\frac{d\hat\sigma_L}{dx}(x,\xi_L)-
\frac{d\hat\sigma_L}{dx}(x,0)\right] ,
\end{eqnarray}
where $\xi_L < 0$ is the position of the Landau pole in the
strong coupling and the function $\Phi(\xi)$ is specified 
in Eq.~(2.25) of \cite{BBB}.
Infrared renormalons correspond to nonanalytic terms in the
expansion of $d\hat\sigma_L/dx (x,\xi) $ at small $\xi$
\begin{eqnarray}\label{expansion}
\frac{d\hat\sigma_L}{dx}(x,\xi) & = & \frac{d\hat\sigma_L}{dx}(x,0)
+ f_1(x)\sqrt{\xi} \nonumber \\
& + & f_2(x)\,\xi\ln\xi 
\end{eqnarray}
and are interpreted as indications of nonperturbative power corrections
of the form
\begin{equation}\label{HT}
\frac{d\hat\sigma_L}{dx} = \frac{d\hat\sigma_L^{\rm pert}}{dx}
-\frac{\mu_{\rm IR}}{Q} f_1(x) - \frac{\mu^2_{\rm IR}}{Q^2} f_2(x) -\ldots 
\end{equation}
Their size can be estimated by the corresponding ambiguity in the summation
of the perturbative series, which is of  order of the imaginary 
part (divided by $\pi$) of the sum in (\ref{BS}). Note that identifying 
the $x$-dependence of the power corrections in (\ref{HT}) with 
the $x$-dependence of the IR renormalon ambiguity or, equivalently, 
the coefficients of non-analytic terms in (\ref{expansion}) is an assumption 
which can not be justified from first principles. Since IR renormalons 
in short-distance quantities are related to ultraviolet ambiguities in 
higher-twist matrix elements, we refer to this assumption as the 
`ultraviolet dominance' of higher-twist corrections. 
  
\section{Perturbative series for $\sigma_L$}

In this section we consider perturbative corrections to $\sigma_L$, 
written as
\begin{equation}
\sigma_L = \sigma_0\,\frac{\alpha_s}{\pi}\,\left[1+
\sum_{n=0}^\infty d_n\,(-\beta_0\alpha_s)^n\right] ,
\end{equation}
where $\sigma_0$ is the Born total $e^+e^-$ cross section. As 
mentioned earlier, we approximate the exact higher-order coefficient 
by its value in the `large-$\beta_0$' limit, where $\beta_0$ is restored 
from the term with the largest power of $N_f$ at each order. This 
approximation, called `naive nonabelianization' in 
\cite{BEN95}, reduces to the familiar BLM prescription for $n=1$. To 
see how it works, we rewrite the exact $\alpha_s^2$ correction in 
(\ref{sigmaL}) as 
\begin{equation}
d_1=6.17-0.7573/(-\beta_0) .
\end{equation}
With $-\beta_0=0.61$ for $N_f=5$, neglecting the second term gives an 
accuracy of about 25\%. We have calculated the coefficients $d_n$ in higher
orders, in the $\overline{\mbox{MS}}$ scheme. The `primary' and 
`secondary' quark contributions, $d_n^{[p]}$ and $d_n^{[s]}$, 
respectively, add to $d_n$ as $d_n=d_n^{[p]}/3+2 d_n^{[s]}/3$. 
A few lower order results up to order $\alpha_s^4$ are
\begin{equation}
\label{eq1}
d_1^{[p]} = 11/2 \quad d_2^{[p]} = 29.8 \quad d_3^{[p]} = 164 ,
\end{equation}
\begin{equation}
\label{eq2}
d_1^{[s]} = 13/2 \quad d_2^{[s]} = 46.0 \quad d_3^{[s]} = 369 .
\end{equation}
The sum of these contributions to all orders is
conveniently written in terms of `enhancement factors'
relative to the leading order contribution \cite{BBB} 
defined by 
\begin{equation}
M^{[p,s]}(\alpha_s) = 1 + \sum_{n=0}^\infty (-\beta_0\alpha_s)^n 
d_n^{[p,s]}~,
\end{equation}
so that
\begin{equation}
 \sigma_L^{({\rm BS})} =\sigma_0 \frac{\alpha_s}{\pi}
\left[\frac{1}{3}M^{[p]}+\frac{2}{3}M^{[s]}\right].
\end{equation}
For various values of $\alpha_s(M_Z)$ we get at $Q=M_Z$
\begin{eqnarray}\label{Mfactors}
\lefteqn{\alpha_s =0.110:}
\nonumber\\
 &M^{[p]}=1.59, & M^{[s]} = 1.92\pm 0.05~.
\nonumber\\  
\lefteqn{\alpha_s =0.120: }
\nonumber\\ 
 &M^{[p]}=1.68,  & M^{[s]} = 2.08\pm 0.08~.
\nonumber\\  
\lefteqn{\alpha_s =0.130:}
\nonumber\\ 
 &M^{[p]}=1.79,  & M^{[s]} = 2.23\pm 0.12~.
\end{eqnarray}
The given numbers correspond to a principal value definition of the 
Borel integral and the uncertainties roughly coincide with the 
size of the minimal term in the series\footnote{The corresponding 
uncertainty for $M^{[p]}$ is small in comparison with the one for 
$M^{[s]}$ and is omitted.}. Let us add the following comments:

(i) The perturbative coefficients in (\ref{eq1}), (\ref{eq2}) grow 
rapidly, especially for the secondary quark contribution. This growth 
is related to an IR renormalon, that indicates a $1/Q^2$ correction 
to primary quark fragmentation and a $1/Q$ correction to 
secondary quark fragmentation, see Sect.~4.

(ii) Even though the $1/Q$ power behaviour indicates much 
larger nonperturbative corrections to $\sigma_L$ as compared to 
$\sigma_{tot}$, the moderate size of the minimal term of the 
perturbative series suggests that these corrections are still 
not large at $Q=M_Z$. The relatively large hadronization correction 
for $\sigma_L$ within the JETSET model applied in \cite{OPAL} 
could thus correspond to higher-order perturbative rather than 
nonperturbative effects.

\setlength{\unitlength}{0.7mm}
\begin{figure}[t]
\vspace{-6.8cm}
\begin{picture}(100,200)(0,1)
\mbox{\epsfxsize7.5cm\epsffile{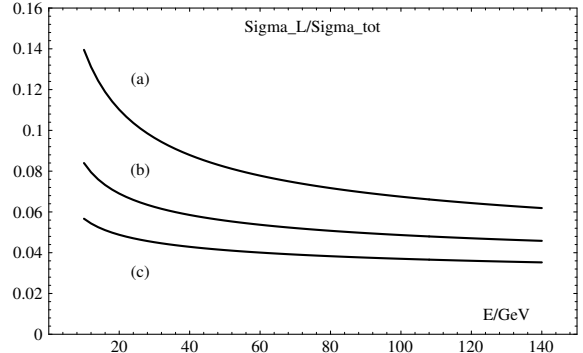}}
\end{picture}
\vspace*{-3.5cm}
\caption{
Longitudinal  fraction in the total
$e^+e^-$ cross section: (c) leading order, (b) next-to-leading order and
(a) resummation of all orders in $\beta_0^n\alpha_s^{n+1}$ corrected for
the exact $O(\alpha_s^2)$ coefficient.}
\end{figure}

(iii) This suggestion is also supported by Fig.~3, where for 
$\alpha_s(M_Z)=0.118$ we have plotted the energy dependence 
of the total longitudinal cross section. Taking into account 
higher-order perturbative corrections [curve (a)] steepens the 
energy dependence, such that it is not far from the JETSET 
prediction, where the steep energy dependence is due to the 
hadronization correction. It is worth noting that the parton 
shower Monte Carlo alone does not yield this energy dependence. 
Experience with similar calculations suggests that the approximation 
of resumming only $(\beta_0\alpha_s)^n$ contributions 
overestimates radiative corrections, so that we expect a more 
realistic estimate in between the curves (a) and (b). 
An exact $O(\alpha_s^3)$ calculation would reduce
the theoretical error considerably. 

(iv) In \cite{DOK95}, universality of the $1/Q$-power correction was 
assumed and a corresponding unique phenomenological parameter 
fitted from the difference between the measured 
average thrust $\langle 1-T\rangle$ and the theoretical second order 
prediction. When added to the second order result (\ref{sigmaL}), 
one obtains a prediction for $\sigma_L$ consistent with data. There 
is no conflict between the procedure of \cite{DOK95} and the one 
presented here, if the phenomenological $1/Q$ correction effectively 
parameterizes the higher-order perturbative contributions added 
in our approach. If universality of power corrections holds, 
these perturbative corrections would also be universal, at least 
asymptotically in large orders. However, from the point of view presented 
here, the universality assumption is not required, since 
higher-order corrections are in principle calculable for each 
observable.

\section{Power corrections}

Returning to (\ref{expansion}), we quote the expansions for the 
invariant mass distributions in (\ref{xi-distributions}):
\begin{eqnarray}
\label{xdist1}
\lefteqn{x\frac{d\hat\sigma_L^{[p]}}{dx}(x,\xi) =} \\ 
\!\!\!\!\!\!&=& \!\!\!\!
\frac{C_F\alpha_s}{2\pi}
\Big\{2x + \xi \ln\xi[8+4\delta(1-x)] + \ldots\Big\}~, \nonumber
\end{eqnarray}
\begin{eqnarray}
\label{xdist2}
\lefteqn{x\frac{d\hat\sigma_L^{[s]}}{dx}(x,\xi)=} \\
\!\!\!\!\!\! &=& \!\!\!\! \frac{C_F\alpha_s}{2\pi}
\Big\{ 4(1-x)(2+2x-x^2)+ 12 x\ln x 
\nonumber\\
& + &\frac{4\xi\ln\xi}{5x^2}\left[
 3+ 30  x - 15 x^3 + 2 x^5+ 15 x^2\ln x\right]
\nonumber\\
& + & \ldots\Big\}~. \nonumber
\end{eqnarray}
Interpreting $\xi$ as $(\Lambda/Q)^2$ where $\Lambda$ is the QCD 
scale, these expressions are valid for $x>\Lambda/Q$. We note 
that for such $x$, all power corrections are at most of order 
$1/Q^2$, in agreement with the result from the light-cone expansion 
of fragmentation processes in \cite{BB91}. We also see that the 
power expansion runs in $\Lambda^2/(Q^2 x)$ for the primary 
quark contribution and $\Lambda^2/(Q^2 x^2)$ for the secondary 
quark (gluon) contribution. The strong divergence of the 
second contribution for small $x$ makes it possible for the moments 
of the $x$-distribution to have parametrically larger power corrections. 
Indeed, we find for the two contributions to the total 
longitudinal cross section 
\begin{eqnarray}
\frac{\hat{\sigma}_L^{[p]}}{\sigma_0}\! \!\!&=&\!\!\! \frac{\alpha_s}{\pi} 
\left[\frac{1}{3} + 0\cdot \sqrt{\xi} + 4\xi\ln \xi + {\cal O}(\xi^2) 
\right] ,
\\
\frac{\hat{\sigma}_L^{[s]}}{\sigma_0}\! \!\!&=& \!\!\!\frac{\alpha_s}{\pi} 
\left[\frac{2}{3} - \frac{5\pi^3}{32}\sqrt{\xi} - 4\xi\ln \xi 
+ {\cal O}(\xi^{\frac{3}{2}}) \right], 
\end{eqnarray}
with a $1/Q$ correction for the secondary quark contribution. Assuming 
ultraviolet dominance of higher-twist corrections, the $x$-distributions 
given in (\ref{xdist1}), (\ref{xdist2}) can be used to model the 
$x$-dependence of power corrections by convoluting the partonic power 
correction with the leading twist fragmentation function \cite{BBM,DAS96}. 
Note that the expressions for the secondary quark contribution differs 
from the gluon contribution to $\sigma_L$ in \cite{DAS96}, because 
the series of higher-order fermion loop diagrams does not reduce to 
the massive gluon calculation performed in \cite{DAS96}. The ensuing 
additional model dependence in the estimate of higher-twist corrections 
will be discussed in \cite{BBM}. Both the calculation here and the 
calculation with a massive gluon coincide in the essential aspects --- 
power corrections of order $\Lambda^2/(Q^2 x^2)$ for finite $x$ and 
$1/Q$ for the integrated longitudinal cross section.

In the remainder of this section, we discuss the origin of the 
$1/Q$ correction in more detail. Adopting for this purpose the massive 
gluon approximation, one finds that the Borel transform for the 
gluon fragmentation contribution factorizes as 
\begin{equation}
B\left[x\frac{d\hat\sigma_L^{[g]}}{dx}\right](x;u) = x^{-2 u} 
\cdot F(u) ,
\end{equation}
when some terms that can not give rise to a $1/Q$ correction are 
omitted. Here $u$ is the Borel parameter and analyticity of 
$F(u)$ for $|u|<1$ corresponds to the statement that only $1/Q^2$ 
corrections arise at finite $x$. Now, the $x$-dependence can be 
absorbed completely into a change of scale in the coupling and 
the Borel integral is given by
\begin{equation}
x\frac{d\hat\sigma_L^{[g]}}{d x} = \int\limits_0^u \!d u\,\exp\left(
-\frac{u}{(-\beta_0\alpha_s(x Q))}\right) F(u).
\end{equation}
The Borel integral (leaving renormalon poles at finite $u$ aside) 
does not exist for $x<\Lambda/Q$, because it diverges at infinity. 
This is a manifestation of the fact that for such small $x$ the 
power expansion breaks down and that power corrections to 
integrated distributions depend sensitively on how the small-$x$ 
region is weighted. Indeed, because of the factorization of the 
$x$-dependence, the $x$-integration is trivial and we get 
\begin{equation}
B\left[\int_0^1 d x x^{1+\gamma} \frac{d\hat\sigma_L^{[g]}}{d x} 
\right] = \frac{F(u)}{1+\gamma-2 u} .
\end{equation}
For the total longitudinal cross section, $\gamma=0$, and the newly 
generated pole at $u=1/2$ corresponds to the $1/Q$ correction discussed 
before. Note that effects due to color coherence and angular ordering 
are expected to change the small-$x$ asymptotic behaviour, which 
could potentially shift the pole to a different value. Clarifying 
the impact of resummation of small-$x$ logarithms requires similar 
efforts to those that have been undertaken to understand the effect 
of Sudakov resummation on power corrections in Drell-Yan production.

Note that the non-uniformity of the power expansion before integration 
over $x$ does not occur for deep inelastic scattering (DIS) processes. 
Given the correspondence of IR renormalons with power ultraviolet 
divergences of higher-twist operators, the difference between 
fragmentation and DIS must be sought in the renormalization properties 
of multi-parton correlation functions that appear in the light-cone 
expansion of \cite{BB91}. For the longitudinal structure function 
in DIS, we find that the quadratic power divergence at one-loop 
of the multi-parton operator
\begin{equation}
g\bar{\psi}(x) \tilde{G}_{\alpha\beta}(v x) x^\alpha \gamma^\beta 
\gamma_5\psi(-x)
\end{equation}
takes the form ($\bar{\alpha}\equiv 1-\alpha$)
\begin{eqnarray}
&&\hspace*{-0.6cm}-\frac{C_F\alpha_s}{4\pi}\frac{\Lambda_{UV}^2}{Q^2}
\!\int\limits_0^1\!d\alpha\,(2-\alpha)\Big\{\bar{\psi}(x)\!\not\! x\psi
(x[\alpha v-\bar{\alpha}])
\nonumber\\
&&\hspace*{0.5cm}+\,\bar{\psi}(x[\alpha v+\bar{\alpha}])
\!\not\! x\psi(-x)\Big\}, 
\end{eqnarray}
that is, the form of a convolution with the leading twist 
contribution. The important point to notice is that the 
operator spreads only a finite distance on the light-cone under 
renormalization. In contrast, the multi-parton correlations that 
appear in fragmentation spread over the entire light-cone 
under renormalization. When the energy fraction $x$ approaches zero, 
the operator becomes sensitive to very large longitudinal distances and 
to how fast the gauge fields decrease at infinity. 
It is this sensitivity to the behaviour at infinity that causes 
a $1/Q$ correction in the longitudinal cross section upon integration 
over $x$. We will return to this point in detail in a future publication 
\cite{BBM}.

\section*{Acknowledgements}

We are grateful to B. Webber for informing us about his related 
work \cite{DAS96} prior to publication. M.~B. and V.~B. thank the 
Institute for Nuclear Theory in Seattle for hospitality during the 
course of this work. This research is supported in part by the 
Department of Energy under contract DE-AC03-76SF-00515.

\end{document}